\documentclass[genTeX]{nrc1}   
\usepackage[pctex32]{graphicx}

 \volyear{99}{2006} \received{2006}
\accepted{2006}
\begin{document}

\title{Sum rules for an atomic hyperfine structure in a magnetic field}
\author{Savely G. Karshenboim}
\address{D. I. Mendeleev Institute for Metrology, St. Petersburg 190005, Russia and
Max-Planck-Institut f\"ur Quantenoptik, Garching 85748, Germany;
e-mail: sek@mpq.mpg.de}

\shortauthor{S. G. Karshenboim}

\maketitle

\begin{abstract}
Sum rules for the energy levels of a hyperfine multiplet in a
constant uniform magnetic field is presented. It is found that for
any values of the electron angular moment and the nuclear spin
there are certain linear combinations of energy levels which do
not depend on the magnetic field and can be used to determine the
unperturbated {\em hfs\/} separation in the presence of perturbing
magnetic field. It is also demonstrated that there are other
linear combinations which are linear with the external magnetic
field and hence can be used to determine bound values of the
electron and nuclear magnetic moments. The accuracy of the
approximation within which the result is valid is also discussed.
\end{abstract}
\begin{resume}
Nous ... French version of abstract (supplied by CJP)
   \traduit
\end{resume}

\section{Introduction}

An essential part of present-day high-precision frequency
measurements is related to experiments performed in the presence
of some residual electric or magnetic field. To reduce the
uncertainty caused by the field, one tries to perform the
measurement on those sublevels which are under control, e.g., for
sublevels slightly affected by the perturbing field. Here, we
study another possible option to deal with a residual magnetic
field. We show that despite the presence of a perturbing magnetic
field shifting the energy levels, there are some special
combinations of energies which do not depend on the magnetic field
at all. In the case of microwave measurements that may be used,
e.g., to determine the hyperfine separation in the ground state.
There are also some more specific combinations, and in particular
ones where the dependence on the magnetic field is linear.

A constant homogenous magnetic field shifts  different sublevels
of hyperfine-structure ({\em hfs\/}) multiplets  differently.
However, as it is well known, a specific combination of energy of
the sublevels of any $nS$ state in the hydrogen atom at presence
of the {\em dc\/} homogenous magnetic field
\begin{equation}\label{brabi}
E^{0}_{\rm hfs}=E^{\bf H}\bigl(1,\,+1\bigr)+E^{\bf
H}\bigl(1,\,-1\bigr) -E^{\bf H}\bigl(1,\,0\bigr)-E^{\bf
H}\bigl(0,\,0\bigr) \label{BrRa}
\end{equation}
remains field-independent. Here: superscript ``${\bf H}$'' stands
for the energy perturbed by the magnetic field ${\bf H}$ and
``${0}$'' is for the unperturbed levels. We use a simplified
notation in which $E^{\bf H}\bigl(1,\,+1\bigr)$ stands for $E^{\bf
H}\bigl(F=1,\,F_z=+1\bigr)$ {\em etc\/}. The direction of the
magnetic field is along the $z$ axis. Here ${\bf F}$ is the
complete atomic angular moment ${\bf F}={\bf J}+{\bf I}$. The
electron angular moment $J$ in the case of the $S$-electron is
$1/2$, and the nuclear magnetic moment $I$ for hydrogen is also
$1/2$. The magnetic quantum number $F_z$ is a well-defined quantum
number in presence of both the external magnetic field and the
hyperfine interaction of the magnetic moment of the nucleus with
the electron angular motion, while $F$ is not in the case when the
Zeeman shifts are comparable with the hyperfine separation. Still
we can mark a level with its value of $F$ at zero magnetic field.
The other quantum numbers (such as the principal quantum number
$n$ and orbital moment $l$ in the case of hydrogen) are not
presented, but indeed it is assumed that all energies are related
to the same {\em hfs\/} multiplet.

The property of the levels in the magnetic field presented in
(\ref{brabi}) has been numerously applied, e.g., to study the
hydrogen \cite{hydr} and muonium \cite{mu} hyperfine structure.

Here we discuss different field-independent combinations of the
energy levels for arbitrary electron states (arbitrary $J$) and
nuclear magnetic moments ($I$). We assume for most of our study
here that any effects due to the nuclear quadrupole electric
moment can be neglected and consider the magnetic field as a weak
one in a sense that the field affects the levels inside the {\em
hfs\/} multiplets but does not mix different multiplets.

We also consider in this paper another kind of specific linear
combinations of energy levels, which depend on a value of the
magnetic field, but their dependence is linear. A ratio of two
linearly-dependent combinations is indeed field-independent. Such
linearly-dependent combinations can be used to determine magnetic
moments of the electron shell and the nucleus. These values
related to the nucleus and the electron(s) in the atom are
somewhat different from their free values, being affected by
binding effects. Comparing various linear combinations, one can
determine a dimensionless ratio of these two magnetic moments.

Measuring various data in presence of a magnetic field, one can in
principle determine a value of the field and thus to introduce
corrections and arrive at unperturbed energy levels and eventually
find magnetic moments. There are two main advantages of the sum
rules derived in our paper.
\begin{itemize}
\item They are derived here for an arbitrary two-spin system
without use of any perturbation techniques and hence the Zeeman
shifts can be of any value compared with the hyperfine splitting.
\item Because of the linearity of the sum rules the combinations
can be easily calculated directly from experimental data and
broadening of lines due to inhomogeneity of the field, which
happens in actual experiments, could be cancelled in part.
\end{itemize}

\section{The Breit-Rabi formula}

The cancellation of the dependence on the magnetic field in
(\ref{BrRa}) has been known as a specific result for the
$nS$-state which originates from the Breit-Rabi formula
\cite{breit,bethe}
\begin{equation}
E^{\bf H}_{\rm magn} \bigl(I\pm1/2,\,F_z\bigr)=-\mu_{\rm nucl}\, H
\pm {\Delta E\over 2}\,\sqrt{1+{4F_z \over 2I+1}\,x+x^2}
\label{BreitRabi}\;,
\end{equation}
where
\begin{equation}
x=x(\Delta E)=\frac{(\mu_J-\mu_{\rm nucl})\,H}{\Delta E}
\end{equation}
and
\begin{equation}\label{deltaE}
\Delta E = E^0\bigl(F=I+1/2\bigr)-E^0\bigl(F=I-1/2\bigr)
\end{equation}
is the {\em hfs\/} separation of the states with $F=I\pm 1/2$ at
zero magnetic field ${\bf H}$. The electron magnetic moment
\begin{equation}
\mbox{\boldmath$\mu$}_J = -\mu_J{{\bf J}\over J}=
-g_J\,\mu_B\,{\bf J}
\end{equation}
depends on the electron state, mainly on values of angular
momentum $J$ and orbital momentum $L$. A slight dependence on
principle quantum number $n$ is to appear from perturbation theory
due to relativistic and QED effects. The nuclear magnetic moment
\begin{equation}
\mbox{\boldmath$\mu$}_{\rm nucl}=\mu_{\rm nucl}{{\bf I}\over
I}=g_N\,\mu_N\,{\bf I}
\end{equation}
also contains a weak dependence on the atomic state via the small
relativistic and QED contributions to $g_N$. The Breit-Rabi
formula is valid for a single electron atom in an $S$ state. Our
approach developed further is valid for an atom with any number of
electrons. Referring to `electron' magnetic moment or `electron'
quantum numbers we will not distinguish between single-electron
values and values for the electron shell.

\begin{figure}[phbt]
\begin{center}
\includegraphics[width=0.45\textwidth]{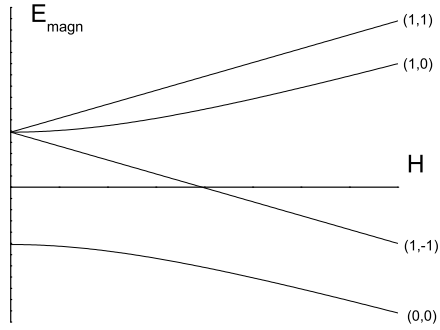}
\end{center}
\caption{Zeeman effect of the $nS$ sublevels in hydrogen (not to
scale). Dependence of the energy of the $^3S_1$ triplet levels and
the $^1S_0$ singlet level on the strength of the constant uniform
field $H$ is plotted.\label{ZeeH}}
\end{figure}

\begin{figure}[phbt]
\begin{center}
\includegraphics[width=0.45\textwidth]{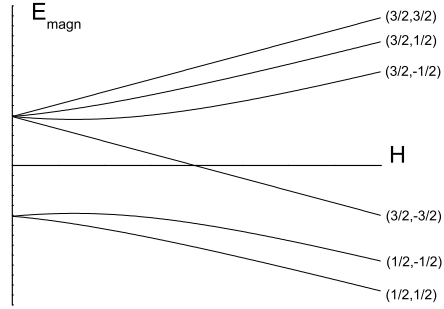}
\end{center}
\caption{Zeeman effect of the $nS$ sublevels in deuterium (not to
scale). \label{ZeeD}}
\end{figure}

Four energy sublevels of an $nS$ state in the hydrogen atom as a
function of the magnetic field are depicted in Fig.~\ref{ZeeH}. A
similar behavior of six hyperfine  sublevels of an $nS$ in
deuterium is plotted in Fig.~\ref{ZeeD}. Let us briefly discuss
our notation. The energy $E^{\bf H}\bigl(F,\,F_Z\bigr)$ is not the
energy of a state with a quantum numbers $F$ and $F_z$ because $F$
is not a well-defined quantum number in presence of the magnetic
field. We consider the energy of a level as a function of the
field strength $H$ and mark it with a value of $F$ at zero field
when $F$ is well defined. We hope this notation is not confusing.

One can conclude from (\ref{BreitRabi}) that there is a specific
combination of few levels for an electronic $nS$ state and any
nuclear spin $I$ which is field-independent. To see that, first we
note that
\begin{equation}\label{maxF}
E^{\bf H}_{\rm magn} \bigl(I+1/2,\,\pm (I+1/2)\bigr)
=-\mu_{\rm nucl} H + {\Delta E\over 2}\bigl(1\pm x\bigr)\;,
\end{equation}
and thus
\begin{equation}\label{maxF-}
E^{\bf H}_{\rm magn} \bigl(I+1/2,\,+(I+1/2)\bigr)+E^{\bf H}_{\rm
magn} \bigl(I+1/2,\,-(I+1/2)\bigr)=-2\mu_{\rm nucl} H + \Delta
E\;.
\end{equation}
Next, we find that
\begin{equation}\label{sqrt_x}
E^{\bf H}_{\rm magn} \bigl(I\pm 1/2,\,+(I- 1/2)\bigr)=-\mu_{\rm
nucl} H \pm {\Delta E\over 2}\sqrt{1+{2(2I-1) \over 2I+1}x+x^2}\;,
\end{equation}
and
\begin{equation}\label{Ilin1}
E^{\bf H}_{\rm magn} \bigl(I+ 1/2,\,+(I- 1/2)\bigr)+E^{\bf H}_{\rm
magn} \bigl(I- 1/2,\,(I- 1/2)\bigr) =-2\mu_{\rm nucl} H\;.
\end{equation}
Similarly, we obtain
\begin{equation}\label{Ilin2}
E^{\bf H}_{\rm magn} \bigl(I+ 1/2,\,-(I- 1/2)\bigr)+E^{\bf H}_{\rm
magn} \bigl(I- 1/2,\,-(I- 1/2)\bigr)=-2\mu_{\rm nucl} H\;.
\end{equation}

Finally we find two field-independent combinations:
\begin{eqnarray}\label{four+}
\Delta E&=&
\Bigl[E^{\bf H}\bigl(I+1/2,\,+(I+1/2)\bigr)+E^{\bf H}\bigl(I+1/2,\,-(I+1/2)\bigr)\Bigr] \nonumber\\
&-& \Bigl[E^{\bf H}_{\rm magn} \bigl(I+ 1/2,\,+(I-
1/2)\bigr)+E^{\bf H}_{\rm magn} \bigl(I- 1/2,\,+(I-
1/2)\bigr)\Bigr]\;,
\end{eqnarray}
and
\begin{eqnarray}\label{four-}
\Delta E&=&
\Bigl[E^{\bf H}_{\rm magn} \bigl(I+1/2,\,+(I+1/2)\bigr)+E^{\bf H}_{\rm magn} \bigl(I+1/2,\,-(I+1/2)\bigr)\Bigr] \nonumber\\
&-& \Bigl[E^{\bf H}_{\rm magn} \bigl(I+ 1/2,\,-(I-
1/2)\bigr)+E^{\bf H}_{\rm magn} \bigl(I- 1/2,\,-(I-
1/2)\bigr)\Bigr]\;,
\end{eqnarray}
where $\Delta E$ is defined by (\ref{deltaE}).

For practical application their symmetric combination can be more
appropriate
\begin{eqnarray}\label{six}
\Delta E&=&
\Bigl[E^{\bf H}_{\rm magn} \bigl(I+1/2,\,+(I+1/2)\bigr)+E^{\bf H}_{\rm magn} \bigl(I+1/2,\,-(I+1/2)\bigr)\Bigr] \nonumber\\
&-&{1\over 2}\biggl\{
\Bigl[E^{\bf H}_{\rm magn} \bigl(I+ 1/2,\,+(I- 1/2)\bigr)+E^{\bf H}_{\rm magn} \bigl(I+ 1/2,\,-(I- 1/2)\bigr)\Bigr] \nonumber\\
&+&\Bigl[E^{\bf H}_{\rm magn} \bigl(I- 1/2,\,+(I-
1/2)\bigr)+E^{\bf H}_{\rm magn} \bigl(I- 1/2,\,-(I-
1/2)\bigr)\Bigr] \biggr\}\;.
\end{eqnarray}

Recent progress in optical measurements offers an opportunity to
determine {\em hfs\/} interval of excited states studying optical
transitions, like, for example, two-photon excitation $1S\to 2S$
in hydrogen and deuterium (see, e.g., \cite{opt}). For such a
measurement the observed lines are to be related to transitions
which conserve $F$ and $F_z$. If the measurement is performed with
a residual field, the sum rules with vanishing field dependence
may be helpful. It is advantageous for practical applications to
express the results in terms of the absolute energy of the levels.
On contrary, the energy in (\ref{BreitRabi}) is a relative energy
defined in such a way that $E\bigl(F=I\pm1/2\bigr)=\pm \Delta E$
at zero magnetic field. The complete energy is
\begin{equation}
E\bigl(\lambda,\; F,\; F_z\bigr) = E_{\rm Coul}\bigl(\lambda\bigr)
+ E^{\bf H}_{\rm magn} \bigl(F,\; F_z\bigr)\;,
\end{equation}
where $\lambda$ stands for quantum numbers which describe the
energy levels neglecting hyperfine interaction and external
magnetic field, i.e. $nlj$ in the case of hydrogen atom. Indeed,
there is still a possibility for an arbitrary additive constant,
but that should be the same for any states. Hopefully, equations
(\ref{four+}), (\ref{four-}) and (\ref{six}) are organized in such
a way that it remains valid after a substitute
\begin{equation}
E^{\bf H}_{\rm magn} \bigl(F,\; F_z\bigr) \to E\bigl(\lambda,\; F,
\; F_z\bigr)\;.
\end{equation}

\section{Atom in a magnetic field: an arbitrary electronic state}

The field-independent results in (\ref{four+}) and (\ref{six})
have been obtained from an explicit expression for the energy
levels (\ref{BreitRabi}) \cite{breit}. Now we are to show that
such a kind of expressions can be derived in a more general case
for an arbitrary electronic state and any value of the nuclear
spin. The only limitation is a complete neglection of the
quadrupole contribution (if it is present), and thus the
separations between levels at zero magnetic field are determined
by a single parameter, which is the nuclear magnetic moment. We
consider an atom in a constant uniform magnetic field which
induces the energy shifts much smaller than the fine structure, so
we can study a reduced Hamiltonian defined as an operator in a
certain spin and angular-momentum space for a particular electron
state
\begin{equation}\label{hMagn}
H_{\rm magn} = - \bigl(\mbox{\boldmath${\mu}$}_J \cdot {\bf
H}\bigr) - \bigl(\mbox{\boldmath${\mu}$}_{\rm nucl}\cdot {\bf
H}\bigr)
 + \bigl({\bf J}\cdot {\bf I}\bigr) A\;,
\end{equation}
where $A$ is the {\em hfs\/} constant which depends on the
electron state. $H_{\rm magn}$ is an operator in the space of the
spin and angular-momentum degrees of freedom. There are two
natural basic sets to describe all such states. One,
$\bigl(F\,,F_z\bigr)$, is appropriate for low magnetic field,
while the other, $\bigl(I_z\,,J_z\bigr)$, is a proper one in the
case when the Zeeman effect dominates over the {\em hfs\/}
separations. An important point is that $F_z$ is a good quantum
number for both sets and
\begin{equation}
\Bigl[F_z,\;H_{\rm magn}\Bigr]=0\;.
\end{equation}

We demonstrate now, that even in the case of arbitrary values of
the nuclear spin and the electron angular moment, a derivation
similar to that for (\ref{four+}) is still possible. It may be
constructed in a similar way with four (six) energy levels. To
show that we remind, that, since $F_z$ is a good quantum number,
it is unnecessary to diagonalize the Hamiltonian $H_{\rm magn}$
for all {\em hfs\/} states simultaneously. One can do that sector
by sector, with each sector being related to a particular value of
$F_z$. There is no interference between sectors different values
of $F_z$.

Let us return to the previously discussed well-known result for an
$nS$ state (see (\ref{four+}) and (\ref{six})). Technically, it is
has been successfully derived because of two basic properties of
Breit-Rabi expression (\ref{BreitRabi}) for energy levels at
magnetic field.
\begin{itemize}
\item There is only a single state with $F_z = +(I+1/2)$ and its
energy has a linear dependence on magnetic field. The same is for
the state with $F_z = -(I+1/2)$. The field-dependence of the
energy is presented in (\ref{maxF}). Since, no re-diagonalization
has been involved the dependence remains linear.
\item There are two states for each other values of $F_z$ and in
particular for $F_z=+(I-1/2)$. To diagonalize the related
Hamiltonian one needs to deal with $2\times2$ matrices only. In a
basis of $(F,\, F_z)$ states, the diagonal matrix elements are
related to the hyperfine interaction, while the off-diagonal
elements are results of interaction with the magnetic field. The
same is true for the substates with $F_z = -(I-1/2)$. The result
of re-diagonalization is given by (\ref{sqrt_x}).
\end{itemize}

Returning to an arbitrary state, we note that there is a certain
similarity. The $F$ value varies between $F_{\rm max}=I+J$ and
$F_{\rm min}=\vert I-J \vert$ with step $\Delta F=1$. The maximal
value of $F_z$ ($F_z=+(I+J)$) is possible only for $F=F_{\rm
max}$. The same is correct for $F_z=-(I+J)$. Since for these
values there is a single hyperfine sublevel for each, the energy
dependence is trivial (cf. (\ref{maxF}))
\begin{equation}\label{arbitrary1}
E^{\bf H}_{\rm magn} \bigl(I+J,\,\pm (I+J)\bigr) =-\mu_{\rm nucl}
H + {\Delta E_1\over 2}\bigl(1\pm x_1\bigr)\;,
\end{equation}
Similarly to the states $F_z =\pm(I+1/2)$ in the previous
consideration, the field dependence is linear. There are two
parameters introduced here: $\Delta E_1= 2E^{0}_{\rm magn}
\bigl(F=I+J\bigr)$ is the energy for the levels with maximal
angular momentum $F=F_{\rm max}$ at zero field, and $x_1=x(\Delta
E_1)$.

The energy $E^{0}_{\rm magn}$ can be in principle defined in an
arbitrary way and as matter of fact it actually cancels out in
$\Delta E_ 1\,x_1$ term in (\ref{arbitrary1}), while any additive
constant has sense only if we discuss a few levels, not a single
one. We postpone any exact definition of $E^{0}_{\rm magn}$ until
it will be necessary for further consideration.

Considering $F_z=F_{\rm max}-1$, we note that there are now two
states with $F=F_{\rm max}$ and $F=F_{\rm max}-1$, which are split
at zero magnetic field. The same situation is indeed for
$F_z=-\vert F_{\rm max}-1 \vert$ That is similar to $F_z =
\pm(I-1/2)$. The energy levels are (cf. (\ref{sqrt_x}))
\begin{equation}\label{arbitrary2}
E^{\bf H}_{\rm magn} \bigl(I+J-1/2\pm 1/2,\,(I+J-
1)\bigr)=\overline{E} -\mu_{\rm nucl} H \pm {\Delta E_2\over
2}\sqrt{1+C(I,J)\cdot x_2+x_2^2}\;.
\end{equation}
Here $\Delta E_2=E^{0}_{\rm magn} \bigl(F=I+J\bigr)-E^{0}_{\rm
magn} \bigl(F=I+J-1\bigr)$ is the {\em hfs\/} splitting between
levels with $F=F_{\rm max}$ and $F=F_{\rm max}-1$;
\[
\overline{E}=\frac{E^{0}_{\rm magn} \bigl(F=I+J\bigr)+E^{0}_{\rm
magn} \bigl(F=I+J-1\bigr)}{2}\;;
\]
and we still do not need to specify $E^{0}_{\rm magn}$. In other
words, it is not important for this part of consideration from
what level we count the energy.

We do not specify also constant $C(I,J)$ in the equation above,
which depends on $I$ and $J$, since that is not necessary for our
further evaluations.

Limiting our consideration by these six states (with $F_z=\pm
F_{\rm max}$ and $F_z=\pm\vert F_{\rm max}-1 \vert$, we easily
derive expressions similar to (\ref{four+}), (\ref{four-}) and
(\ref{six}).

Firstly, we note that
\[
\Delta E_1 \cdot x_1 = \Delta E_2 \cdot x_2\;.
\]
Secondly, we choose such a definition of the energy for a moment,
that satisfies the condition
\begin{equation}\label{defE0}
\overline{E}=0\;,
\end{equation}
which is a way to fix an additive constant in definition of
energy. Once we do so, the equations (\ref{arbitrary1}) and
(\ref{arbitrary2}) take now the form (cf. (\ref{maxF}) and
(\ref{maxF-}))
\begin{eqnarray}
E^{\bf H}_{\rm magn} \bigl(I+J,\,\pm (I+J)\bigr) &=&-\mu_{\rm
nucl}
H + {\Delta E_ 2\over 2}\bigl(1\pm x_2\bigr)\;,\nonumber\\
E^{\bf H}_{\rm magn} \bigl(I+J-1/2\pm 1/2,\,(I+J- 1)\bigr)&=&
-\mu_{\rm nucl} H \pm {\Delta E_2\over 2}\sqrt{1+C(I,J)\cdot
x_2+x_2^2}\;,\nonumber
\end{eqnarray}
and the result
\begin{eqnarray}
\Delta E_2 &=&  \Bigl[E^{\bf H}_{\rm magn} \bigl(I+J,\,+(I+J)\bigr)+E^{\bf H}_{\rm magn} \bigl(I+J,\,-(I+J)\bigr)\Bigr] \nonumber\\
&-&{1\over 2}\biggl\{
\Bigl[E^{\bf H}_{\rm magn} \bigl(I+J,\,+(J+I-1)\bigr)+E^{\bf H}_{\rm magn} \bigl(F=I+J,\,-(I+J-1)\bigr)\Bigr] \nonumber\\
&+&\Bigl[E^{\bf H}_{\rm magn} \bigl(I+J-1,\,+(I+J-1)\bigr)+E^{\bf
H}_{\rm magn}
\bigl(I+J-1,\,-(I+J-1)\bigr)\Bigr]\biggr\}\label{arbitrary6}
\end{eqnarray}
becomes obvious.

Above we have defined the energy by (\ref{defE0}); however, we
note that (\ref{arbitrary6}) (cf. (\ref{six})) allows a
redefinition of energy by introducing an additive constant. That
reads that we are free to change definition of energy with an
additive constant if necessary and this constant can depend on any
quantum numbers not involved into hyperfine interaction.

\section{Sum rules: field-independent combinations \label{sr0}}

Above we have taken advantage of similarity in certain expressions
for simple previously known case and for a general situation. The
similarity is related to sectors determined by a value of $F_z$
which have one or two substates only. That is the case of two
maximal possible values of $F_z$ (and the opposite values with
negative $F_z$). Any other sector with a lower value of $\vert F_z
\vert$ involves more than two states and any explicit expression
for energy becomes more complicated. It is even unclear whether it
is possible to obtain such expressions. Below we develop an
alternative approach to derive various field-independent values as
linear combinations of energies in a constant uniform magnetic
field. The approach does not need any explicit expressions for
energy levels at magnetic field.

First, we find a trace of the magnetic Hamiltonian
\begin{equation}
S_0={\rm Sp}\,\Bigl\{H_{\rm magn}\Bigr\}=\sum_\zeta \langle \zeta
\vert H_{\rm magn}\vert \zeta\rangle\;,
\end{equation}
where the summation is performed over a complete basis set of
sublevels, $\zeta$, of the {\em hfs\/} multiplet. There are two
basis sets useful for calculations: one is related to states with
a fixed values of ${\bf F}^2$ and $F_z$ and the other is for the
states with determined values of $J_z$ and $I_z$. Since the
summation is over a complete set the result does not depend on the
choice between these two sets.

It is obviously that
\begin{equation}\label{S00}
S_0={\rm Sp}\,\Bigl\{H_{\rm magn}\Bigr\}=0\;.
\end{equation}

A more general trace of interest is
\begin{equation}\label{s2n}
S_{2n}={\rm Sp}\,\Bigl\{F_z^{2n} \,H_{\rm magn}\Bigr\}\;.
\end{equation}
Still, the field-dependent part obviously vanishes because
\begin{equation}
{\rm Sp}\,\Bigl\{F_z^{2n} \,\mbox{\boldmath${\mu}$}_J \cdot {\bf
H}\Bigr\}={\rm Sp}\,\Bigl\{F_z^{2n} \,\mbox{\boldmath${\mu}$}_{\rm
nucl} \cdot {\bf H}\Bigr\}=0\;.
\end{equation}
The remaining part is field-independent
\begin{equation}\label{sum2na}
S_{2n}={\rm Sp}\,\Bigl\{F_z^{2n} \,\bigl({\bf J}\cdot {\bf
I}\bigr)\Bigr\}\,A ={1\over2}\sum_{F,\,F_z}\Bigl\{F_z^{2n}
\,\bigl[F(F+1)-I(I+1)-J(J+1)\bigr]\Bigr\}\,A\;.
\end{equation}
The {\em hfs\/} constant $A$ determines all energy separations
inside the {\em hfs\/} multiplet and e.g. one can find
\begin{equation}
E^0_{\rm magn} \bigl(F=I+J\bigr)-E^0_{\rm magn} \bigl(F=|I-J|\bigr)={1\over2}\Bigl[\bigl(I+J\bigr)\bigl(I+J+1\bigr)-
\bigl(|I-J|\bigr)\bigl(|I-J|+1\bigr)\Bigr]\,A\;.
\end{equation}

Let us discuss values which can be measured directly. The complete Hamiltonian is of the form
\begin{equation}
H_{\rm tot} = H_0+H_{\rm magn}\;,
\end{equation}
where the first term depends on the electronic state and does not
depend on magnetic parameters ${\bf I}$ and ${\bf J}$, while the
second term depends on them only, but not on detail of the atomic
state. One can see that
\begin{equation}\label{summm}
{\rm Sp}\,\Bigl\{F_z^{2n}\,\bigl(H_0+H_{\rm magn}\bigr) -
\overline{F_z^{2n}} \bigl(H_0+H_{\rm magn}\bigr)\Bigr\} =S_{2n}+
H_0\,{\rm Sp}\,(F_z^{2n} - \overline{F_z^{2n}})-
\overline{F_z^2}\,S_0 =S_{2n}\;,
\end{equation}
where $\overline{F_z^{2n}}$ is an average value over all the states
\begin{equation}
\overline{F_z^{2n}}=\frac{{\rm
Sp}\,\Bigl\{F_z^{2n}\Bigr\}}{(2J+1)(2I+1)}\;.
\end{equation}
After some transformations we arrive to a final field-independent
sum rule
\begin{eqnarray}\label{finals2n}
S_{2n}=\sum_{F,F_z}{}\Bigl(F_z^2-\overline{F_z^{2n}}\Bigr)E^{\bf
H}\bigl(F,F_z\bigr)&=&
{1\over2}\sum_{F,F_z}{}\Bigl\{F_z^{2n} \,\bigl(F(F+1)-I(I+1)-J(J+1)\bigr)\Bigr\}\nonumber\\
&\times&\frac{E^0_{\rm magn} \bigl(F=I+J\bigr)-E^0_{\rm magn} \bigl(|I-J|\bigr)}{(I+J+1)(I+J)-(|I-J|+1)|I-J|}
\;,
\end{eqnarray}
where the sum is over all spin states. The equation for $S_2$ in
the case of hydrogen reproduces (\ref{BrRa}) and in the case of
deuterium (\ref{six}).

\section{Sum rules: combinations linear in magnetic field \label{srH}}

The relation (\ref{finals2n}), valid for an arbitrary value of
$n$, allows to obtain an infinite number of relationships.
Obviously, not all of them are independent\footnote{We understand
independent relations as relations which present independent
linear combinations of $E^{\bf H}_{\rm magn} \bigl(F,\,F_z\bigr)$.
Indeed, they are not independent in a sense that all are
eventually expressed in terms of the hyperfine constant $A$.} of
each other. Let us discuss a number of independent identities. The
equation, presented above, combines the values
\begin{equation}
\widetilde{E}^{\bf H}_{\rm magn} (F_z) =\sum_F\Bigl\{E^{\bf
H}_{\rm magn} \bigl(F,+F_z\bigr)+E^{\bf H}_{\rm magn}
\bigl(F,-F_z\bigr)\Bigr\}
\end{equation}
for not-negative $F_z$. The number of independent identities,
derived in such a way, cannot exceed the number of different
values of $\widetilde{E}^{\bf H}(F_z)$. The latter is equal to
$I+J+1$ for integer $I+J$ and $I+J+1/2$ for semi-integer. We note
that all identities allow free additive normalization of energy.
It means that we, e.g., should consider $\widetilde{E}^{\bf
H}_{\rm magn} (F_z)-\widetilde{E}^{\bf H}_{\rm magn} (0)$ rather
than $\widetilde{E}^{\bf H}_{\rm magn} (F_z)$, which reduces
number of independent sum rules by unity.

Comparing with consideration for $nS$ states, we note that the
field-independent combination is not necessarily a symmetrical sum
of the contributions with $F_z$ and $-F_z$ (cf. (\ref{four+}) and
(\ref{four-})). Meantime, we derived above only symmetric
combinations. Asymmetric field-independent combinations cannot be
found this way.

Nevertheless, we can study odd sum rules related to the properties
of
\begin{equation}\label{s2n1}
S_{2n+1}={\rm Sp}\,\Bigl\{F_z^{2n+1} \,H_{\rm magn}\Bigr\}\;.
\end{equation}
In contrast to the even sum rules (\ref{s2n}), the part related to
the hyperfine interaction of the angular momentum and the nuclear
magnetic moment vanishes
\begin{equation}
{\rm Sp}\,\Bigl\{F_z^{2n} \,\bigl({\bf J}\cdot {\bf
I}\bigr)\Bigr\}\,A =0\;,
\end{equation}
and the rest depends on the magnetic field. The dependence is
quite simple:
\begin{equation}
S_{2n+1}=\sum_{J_z,I_z}\bigl(J_z+I_z\bigr)^{2n+1}\Bigl\{
-g_j\,\mu_B\,J_z+g_N\,\mu_N\,I_z\Bigr\}\,H\,.
\end{equation}
It is linear in the magnetic field.

In such a way we have derived not more than $I+J$ values for
integer $I+J$ and $I+J+1/2$ values of $S_{2n+1}$ for semi-integer
$I+J$, which are linear in magnetic field. Any ratio of them is
indeed a field-independent quantity. That provides us with another
set of values, which cannot be affected by the magnetic field. So,
we derived not more than $I+J-1$ and $I+J-1/2$ field-independent
values independent from each other.

We note that
\begin{equation}
\sum_{J_z,I_z}\bigl(J_z+I_z\bigr)^{2n+1}=0
\end{equation}
and thus
\begin{equation}
S_{2n+1}=\sum_{J_z,I_z}\bigl(J_z+I_z\bigr)^{2n+1}\,E^{\bf H}_{\rm
magn} \bigl(J_z,\,I_z\bigr)\;,
\end{equation}
where we changed the notation and $E^{\bf H}\bigl(J_z,\,I_z\bigr)$
stands for the energy of the levels which for the strong magnetic
field is characterized by eigenvalues of $J_z$ and $I_z$.

\section{Summary}

In our evaluation above we marked the levels with quantum numbers
($F$, $F_z$) or ($I_z$, $J_z$), which both are not good at the
presence at the same time of two effects such as the hyperfine
interaction and interaction with the magnetic field. The indeces
$F$, $I_z$ and $J_z$ corresponded not to well-defined quantum
numbers, but were just indexes related to some limits (`weak' or
`strong' magnetic field). In the case of any formal problem of
derivation of our results for $S_{2n}$ and $S_{2n+1}$ we note that
all of them can be presented as a double sum, when the internal
sum is a sum of all states, $\zeta(F_z)$, with a fixed value of
$F_z$
\begin{equation}
\overline{E} (F_z)=\sum_{\zeta(F_z)}\,E^{\bf H}_{\rm magn} \bigl(\zeta(F_z)\bigr)\;,
\end{equation}
while the external sum is over $F_z$ which is a good quantum
number. In such a way we need only somehow to identify levels and
know their eigennumbers for $F_z$.

Altogether we derived not more than $2(I+J)-1$ field-independent
values (as $S_{2n}$ or as ratio of two $S_{2n+1}$). The sum rules,
derived here, (for $I\geq1$ and $J\geq1$) are not covered by the
Breit-Rabi formula. We also suggested an analog of the Breit-Rabi
formula for six specific levels. The derivation above is valid for
any spin hamiltonian when we can neglect interaction between
different multiplets.

In conclusion, let us discuss two problems related to possible
corrections to the sum rules due to
\begin{itemize}
\item effects of the nuclear structure;
\item quantum electrodynamics effects and atomic structure.
\end{itemize}

We note that the nuclear electric quadrupole moment (or higher
order electric and magnetic moments) affects the identities above
but still the field-dependent combinations exist. Since $F_z$ is a
good quantum number, the sums $S_{2n}$ are field-independent and
calculating all of them one can find contribution of all moments
separately. For instance, the nuclear quadrupole moment
contributes to hyperfine structure if $I\geq1$ and $J\geq1$. The
magnetic Hamiltonian becomes of the form
\begin{equation}
H_{\rm magn} = - \bigl(\mbox{\boldmath${\mu}$}_J \cdot {\bf
H}\bigr) - \bigl(\mbox{\boldmath${\mu}$}_{\rm nucl}\cdot {\bf
H}\bigr)
 + \bigl({\bf J}\cdot {\bf I}\bigr)\,A + 4\bigl({\bf J}\cdot {\bf I}\bigr)^2\,B \;,
\end{equation}
where the constant $B$ is directly related to the quadrupole
contribution to the hyperfine structure. In such a case, the sums
$S_{2n}$ remains field-independent, but they are expressed
linearly in terms of $A$ and $B$ (cf. (\ref{sum2na})). It is
enough to find two field-independent values, e.g., $S_2$ and
$S_4$, to determine two contributions to the hyperfine structure
separately: one caused by the dipole magnetic moment and the other
caused by the electric quadrupole moment, i.e., by $A$ and $B$. We
do not present any explicit relationships but it is clear that
they can be presented if it would be really necessary for any
particular applications. The sum rules offer an accurate method to
compare the magnetic dipole moment and electric quadrupole moment
in the presence of a magnetic field. It can be also used to
compare magnetic moments of the nucleus and an electron by
studying a ratio of linear combinations. If one measures magnetic
field, the linear combinations can be helpful as well (cf., e.g.,
with a high-precision determination of the muonium magnetic moment
in Ref~\cite{mu}).

There is only a source of corrections to the derived sum rules due
to other levels. If the magnetic shift of levels of fine and gross
structure is much smaller than the distance between those levels,
we can separate the levels in theoretical consideration to study
their magnetic structure. If the shift is big, a number of levels
are mixed (cf. the Paschen-Back effect) and our evaluation is
incorrect. It also leads to a Hamiltonian non-linear in the
magnetic field. Indeed, in a low field such effects produce
certain corrections to the Hamiltonian (\ref{hMagn}).
Additionally, these effects produce a shift of the center of
gravity of magnetic multiplet. However, as long as these effects
are small they can be taken into account perturbatively as effects
in the second order in the magnetic field.

All other effects, such as effects of quantum electrodynamics, can
only affect the values of the electron and nuclear magnetic
moments which become slightly different from the free values and
thus introduce some corrections to the energy levels in the
magnetic field. However, the structure of the Hamiltonian
(\ref{hMagn}) is the same and sum rules for the field-independent
contributions are not affected.

\section*{Acknowledgments}

The work was supported in part by RFBR (under grant \#
06-02-16156) and by DFG (under grant \# DFG GZ: 436 RUS
113/769/0). I am grateful to Gordon Drake and Vladimir Ivanov for
useful discussions.

\end{document}